\documentclass[12pt]{iopart}%
\begin{document}
\title[The Cosmological Constant]{Naked Singularity in a Modified Gravity Theory}
\author{Remo Garattini}

\begin{abstract}
The cosmological constant induced by quantum fluctuation of the graviton on a
given background is considered as a tool for building a spectrum of different
geometries. In particular, we apply the method to the Schwarzschild background
with positive and negative mass parameter. In this way, we put on the same
level of comparison the related naked singularity $\left(  -M\right)  $ and
the positive mass wormhole. We discuss how to extract information in the
context of a $f\left(  R\right)  $ theory. We use the Wheeler-De Witt equation
as a basic equation to perform such an analysis regarded as a Sturm-Liouville
problem . The application of the same procedure used for the ordinary theory,
namely $f\left(  R\right)  =R$, reveals that to this approximation level, it
is not possible to classify the Schwarzschild and its naked partner into a
geometry spectrum.

\end{abstract}

\address{Universit\`{a} degli Studi di Bergamo, Facolt\`{a} di Ingegneria,\\ Viale
Marconi 5, 24044 Dalmine (Bergamo) ITALY.\\INFN - sezione di Milano, Via Celoria 16, Milan, Italy}
\ead{remo.garattini@unibg.it}

\section{Introduction}

Almost ten years ago, observation about type I supernovae data revealed that
the Universe is in an acceleration phase\cite{Obs}. Since then, no
satisfactory explanation has been given. Indeed if the
Friedmann-Robertson-Walker model of the universe, based on the Einstein's
field equations is correct, the explanation of such an expansion should be due
to approximately a 76\% of what is known as \textit{Dark Energy}. Dark Energy
is based on the following equation of state $P=-\rho$ (where $P$ and $\rho$
are the pressure pf the fluid and the energy density, respectively). Dark
Energy changes into \textit{Phantom Energy} when $P<-\rho$. Nevertheless,
neither Dark Energy nor Phantom Energy models appear to be satisfactory. In
this scenario, the idea that General Relativity could be modified in something
more general has been considered in recent years. In particular, the
Einstein-Hilbert action $\left(  \kappa=8\pi G\right)  $%
\begin{equation}
S=\frac{1}{2\kappa}\int d^{4}x\sqrt{-g}R+S^{matter}%
\end{equation}
is replaced by\cite{f(R)}%
\begin{equation}
S=\frac{1}{2\kappa}\int d^{4}x\sqrt{-g}f\left(  R\right)  +S^{matter}.
\end{equation}
It is clear that other more complicated choices could be done in place of
$f\left(  R\right)  \cite{CF}.$ In particular, one could consider $f\left(
R,R_{\mu\nu}R^{\mu\nu},R_{\alpha\beta\gamma\delta}R^{\alpha\beta\gamma\delta
},\ldots\right)  $ or $f\left(  R,G\right)  $ where $G$ is the Gauss-Bonnet
invariant or any combination of these quantities\footnote{For a recent riview
on $f\left(  R\right)  $, see Refs\cite{Faraoni,CF}, while a recent review on
the problem of $f\left(  G\right)  $ and $f\left(  R,G\right)  $ can be found
in\cite{NO}.}. Note that in principle one could consider the replacement
$R-2\Lambda_{c}\rightarrow f\left(  R\right)  $ in such a way to avoid the use
of the cosmological constant $\Lambda_{c}$. It is well known indeed, that
there exists a factor of 120 orders of magnitude of discrepancy between the
observed and the computed value. This huge disagreement is known as
\textit{cosmological constant problem}. It is important to remark that the
cosmological constant plays an alternative r\^{o}le with respect to Dark
Energy and modified gravity theories in explaining the acceleration of the
Universe. Nevertheless, nothing forbids to consider them together in the
context of the Wheeler-DeWitt equation (WDW)\cite{De Witt}%
\begin{equation}
\mathcal{H}\Psi\mathcal{=}0,\label{WDW}%
\end{equation}
where\footnote{See Ref.\cite{Remo} for more details.}%
\begin{equation}
\mathcal{H=}\left(  2\kappa\right)  G_{ijkl}\pi^{ij}\pi^{kl}-\frac{\sqrt{g}%
}{2\kappa}\!{}\!\left(  \,^{3}R-2\Lambda_{c}\right)  ,
\end{equation}
$^{3}R$ is the scalar curvature in three dimensions and $G_{ijkl}$ is called
the super-metric. $\mathcal{H}$ represents the time-time component of the
Einstein's field equations without matter fields. It represents a constraint
at the classical level and the invariance under \textit{time}
reparametrization. One can formally re-write the WDW equation as an eigenvalue
problem%
\begin{equation}
\frac{1}{V}\frac{\int\mathcal{D}\left[  g_{ij}\right]  \Psi^{\ast}\left[
g_{ij}\right]  \int_{\Sigma}d^{3}x\hat{\Lambda}_{\Sigma}\Psi\left[
g_{ij}\right]  }{\int\mathcal{D}\left[  g_{ij}\right]  \Psi^{\ast}\left[
g_{ij}\right]  \Psi\left[  g_{ij}\right]  }=\frac{1}{V}\frac{\left\langle
\Psi\left\vert \int_{\Sigma}d^{3}x\hat{\Lambda}_{\Sigma}\right\vert
\Psi\right\rangle }{\left\langle \Psi|\Psi\right\rangle }=-\frac{\Lambda_{c}%
}{\kappa},\label{WDW2}%
\end{equation}
where%
\begin{equation}
\hat{\Lambda}_{\Sigma}=\left(  2\kappa\right)  G_{ijkl}\pi^{ij}\pi^{kl}%
-\frac{\sqrt{g}}{2\kappa}\!{}\,^{3}R
\end{equation}
and%
\begin{equation}
V=\int_{\Sigma}d^{3}x\sqrt{g}.
\end{equation}
It is clear that, what we interpret as an eigenvalue is an induced
cosmological constant and, as pointed out in Ref.\cite{Remo1}, we can use such
an eigenvalue evaluated to one loop in different backgrounds as a tool to
compute a geometrical \textquotedblleft\textit{spectrum}\textquotedblright%
\ based on Zero Point Energy (ZPE). In particular, in Ref.\cite{Remo1},we have
computed the induced cosmological constant generated by a naked singularity
associated to the Schwarzschild metric. The reason for such a choice is
simple: if the Schwarzschild solution is%
\begin{equation}
ds^{2}=-\left(  1-\frac{2MG}{r}\right)  dt^{2}+\left(  1-\frac{2MG}{r}\right)
^{-1}dr^{2}+r^{2}\left(  d\theta^{2}+\sin^{2}\theta d\phi^{2}\right)
,\label{Mp}%
\end{equation}
replacing $M$ with $-M=\bar{M}$, we obtain a naked singularity with the
following line element%
\begin{equation}
ds^{2}=-\left(  1+\frac{2\bar{M}G}{r}\right)  dt^{2}+\left(  1+\frac{2\bar
{M}G}{r}\right)  ^{-1}dr^{2}+r^{2}\left(  d\theta^{2}+\sin^{2}\theta d\phi
^{2}\right)  .\label{MpN}%
\end{equation}
It is immediate to see there is no horizon protecting the singularity. An
immediate consequence of a negative Schwarzschild mass is that if one were to
place two bodies initially at rest, one with a negative mass and the other
with a positive mass, both will accelerate in the same direction going from
the negative mass to the positive one. Furthermore, if the two masses are of
the same magnitude, they will uniformly accelerate forever\footnote{See
Ref.\cite{Bondi}, for negative mass analysis in General Relativity. See also
the problem of the \textit{Cosmic Censorship Conjecture} postulated by R.
Penrose\cite{Penrose}}. Another reason to consider a naked singularity
described by the line element \ref{MpN} is that it naturally represents a form
of \textquotedblleft\textit{Dark Energy}\textquotedblright\ and therefore it
deserves attention. In this paper we would like to extend the analysis of
Ref.\cite{Remo1} to $f\left(  R\right)  $ theories. Although the subject of
this investigation is quite delicate, because as far as we know the subject of
$f\left(  R\right)  $ theories combined with naked singularities has not been
considered even at the classical level, it seems quite reasonable to apply the
scheme of Eq.$\left(  \ref{WDW2}\right)  $ to the negative Schwarzschild mass.
Even in this case, we exclude the contribution of matter fields and the final
result will be due only to quantum fluctuations. In practice, we desire to
compute%
\begin{equation}
\Delta\Lambda_{c}=\Lambda_{c}^{S}-\Lambda_{c}^{N}\geq\left(  \leq\right)  \ 0,
\end{equation}
where $\Lambda_{c}^{S,N}$ are the induced cosmological constant computed in
the different backgrounds. Moreover, the Schwarzschild solution for both
masses, namely $\pm M$ is asymptotically flat. Therefore we are comparing
backgrounds with the same asymptotically behavior. Nevertheless, in
Eq.$\left(  \ref{WDW}\right)  $, surface terms never come into play because
$\mathcal{H}$ as well as $\Lambda_{c}/\kappa$ are energy densities and surface
terms are related to the energy (e.g. ADM mass) and not to the energy density.
We want to point up that even in the case of $f\left(  R\right)  $ theories,
we are neither discussing the problem of forming the naked singularity nor a
transition during a gravitational collapse. Rather the singularity is
considered already existing. The semi-classical procedure followed in this
work relies heavily on the formalism outlined in Refs.\cite{Remo1,CG}.

\section{Positive and negative Schwarzschild mass in a $f\left(  R\right)  =R$
theory}

The Schwarzschild background is simply described by Eq.$\left(  \ref{Mp}%
\right)  $. In terms of the induced cosmological constant of Eq.$\left(
\ref{WDW2}\right)  $, we get%
\begin{equation}
\frac{\Lambda_{0,c}\left(  \mu_{0},r\right)  }{8\pi G}=\frac{1}{64\pi^{2}}%
\sum_{i=1}^{2}\left(  \frac{3MG}{r^{3}}\right)  ^{2}\ln\left(  \left\vert
\frac{4r^{3}\mu_{0}^{2}}{3MG\sqrt{e}}\right\vert \right)  , \label{lambda0mu}%
\end{equation}
where we have removed the ultraviolet divergence with the help of a zeta
function regularization and applied a Renormalization Group equation in order
to avoid a dependence on the mass scale $\mu$.

We know that an extremum appears, maximizing the induced cosmological constant
for%
\begin{equation}
\frac{3MG\sqrt{e}}{4r^{3}\mu_{0}^{2}}=\frac{1}{\sqrt{e}} \label{sol}%
\end{equation}
and leading to%
\begin{equation}
\frac{\Lambda_{0,c}\left(  \mu_{0},r\right)  }{8\pi G}=\frac{\mu_{0}^{4}%
}{4e^{2}\pi^{2}} \label{lambda0mue}%
\end{equation}
or\footnote{See Refs.\cite{Remo,Remo1} for technical details concerning the
reasons of why $r\in\left[  r_{t},\frac{5}{4}r_{t}\right]  $.}%
\begin{equation}
\frac{\Lambda_{0,c}\left(  \mu_{0},r\right)  }{8\pi G}=\left(  \frac
{3MG}{r^{3}}\right)  ^{2}\frac{1}{64\pi^{2}}\qquad r\in\left[  r_{t},\frac
{5}{4}r_{t}\right]  .
\end{equation}
Therefore, it appears that there exists a bound for $\Lambda_{0,c}$%
\begin{equation}
\frac{9}{256\pi^{2}r_{t}^{4}}\leq\frac{\Lambda_{0,c}\left(  \mu_{0},r\right)
}{8\pi G}\leq\frac{225}{4096\pi^{2}r_{t}^{4}}.
\end{equation}
When we consider the naked Schwarzschild metric, we obtain an induced
cosmological constant of the form%
\[
\frac{\Lambda_{0,c}^{naked}\left(  \mu_{0},r\right)  }{8\pi G}%
\]%
\begin{equation}
=\frac{1}{64\pi^{2}}\left[  \left(  \frac{15\bar{M}G}{r^{3}}\right)  ^{2}%
\ln\left(  \left\vert \frac{4r^{3}\mu_{0}^{2}}{15\bar{M}G\sqrt{e}}\right\vert
\right)  +\left(  \frac{9\bar{M}G}{r^{3}}\right)  ^{2}\ln\left(  \left\vert
\frac{4r^{3}\mu_{0}^{2}}{9\bar{M}G\sqrt{e}}\right\vert \right)  \right]  .
\label{Lambda0cN}%
\end{equation}
In order to find an extremum, it is convenient to define the following
dimensionless quantity%
\begin{equation}
\frac{9\bar{M}G\sqrt{e}}{4r^{3}\mu_{0}^{2}}=x, \label{dim}%
\end{equation}
then Eq.$\left(  \ref{Lambda0cN}\right)  $ becomes%
\begin{equation}
\frac{\Lambda_{0,c}^{naked}\left(  \mu_{0},r\right)  }{8\pi G}=-\frac{\mu
_{0}^{4}}{4e\pi^{2}}\left[  x^{2}\ln x+\frac{25}{9}x^{2}\ln\left(  \frac
{5x}{3}\right)  \right]  .
\end{equation}
We find a solution when%
\begin{equation}
\bar{x}=\frac{1}{\sqrt{e}}\left(  \frac{3}{5}\right)  ^{\frac{25}{34}}%
\simeq0.417 \label{sol1}%
\end{equation}
corresponding to a value of%
\begin{equation}
\frac{\Lambda_{0,c}^{naked}\left(  \mu_{0},r\right)  }{8\pi G}=\frac{\mu
_{0}^{4}}{4e^{2}\pi^{2}}\frac{17}{75}5^{\left(  \frac{9}{17}\right)
}3^{\left(  \frac{8}{17}\right)  }\simeq0.328\frac{\mu_{0}^{4}}{4e^{2}\pi^{2}%
}=0.328\frac{\Lambda_{0,c}\left(  \mu_{0},r\right)  }{8\pi G}.
\end{equation}
This means that%
\begin{equation}
\frac{\Lambda_{0,c}^{naked}\left(  \mu_{0},r\right)  }{\Lambda_{0,c}\left(
\mu_{0},r\right)  }=0.328<1. \label{rap}%
\end{equation}
A comment to this inequality is in order. Eq.$\left(  \ref{rap}\right)  $
shows that the Schwarzschild naked singularity has a lower value of ZPE
compared to the positive Schwarzschild mass. This means that, even if the
order of magnitude is practically the same, the naked singularity is less
favored with respect to the Schwarzschild wormhole. We now try to apply the
same method to a modified gravity theory of the form $f\left(  R\right)  $.

\section{Positive and negative Schwarzschild mass one loop energy for a
generic $f\left(  R\right)  $ theory}

In this section, we report the main steps discussed in Ref.\cite{CG} for a
$f\left(  R\right)  $ theory in connection with the Sturm-Liouville problem of
Eq.$\left(  \ref{WDW2}\right)  $. Although a $f\left(  R\right)  $ theory does
not need a cosmological constant, rather it should explain it, we shall
consider the following Lagrangian density describing a generic $f(R)$ theory
of gravity
\begin{equation}
\mathcal{L}=\sqrt{-g}\left(  f\left(  R\right)  -2\Lambda\right)
,\qquad{with}\;f^{\prime\prime}\neq0,\label{lag}%
\end{equation}
where $f\left(  R\right)  $ is an arbitrary smooth function of the scalar
curvature and primes denote differentiation with respect to the scalar
curvature. A cosmological term is added also in this case for the sake of
generality, because in any case, Eq.$\left(  \ref{lag}\right)  $ represents
the most general lagrangian to examine. Obviously $f^{\prime\prime}=0$
corresponds to GR. The generalized Hamiltonian density for the $f\left(
R\right)  $ theory assumes the form\footnote{See Refs.\cite{Querella,CG} for
technical details.}%
\[
\mathcal{H}=f^{\prime}\left(  R\right)  \left[  \left(  2\kappa\right)
G_{ijkl}\pi^{ij}\pi^{kl}{}-\frac{\sqrt{g}}{2\kappa}{}\left(  ^{\left(
3\right)  }R-2\Lambda_{c}\right)  \right]
\]%
\begin{equation}
+2\left(  2\kappa\right)  \left[  G_{ijkl}\pi^{ij}\pi^{kl}+\frac{\pi^{2}}%
{4}\right]  \left(  f^{\prime}\left(  R\right)  -1\right)  +\frac{1}{2\kappa
}\left[  V(\mathcal{P})+2g^{ij}\left(  \sqrt{g}f^{\prime}\left(  R\right)
\right)  _{\mid ij}\right]  .\label{Hamf(R)_1}%
\end{equation}
where%
\begin{equation}
\mathcal{P=}-6\sqrt{g}f^{\prime}\left(  R\right)
\end{equation}
and%
\begin{equation}
V(\mathcal{P})=\sqrt{g}\left[  Rf^{\prime}\left(  R\right)  -f\left(
R\right)  \right]  .\label{V(P)}%
\end{equation}
When $f\left(  R\right)  =R$, $V(\mathcal{P})=0$ as it should be. By imposing
the Hamiltonian constraint and integrating over the hypersurface $\Sigma$, we
obtain%
\[
\int_{\Sigma}d^{3}x\left\{  \left[  \left(  2\kappa\right)  G_{ijkl}\pi
^{ij}\pi^{kl}{}-\frac{\sqrt{g}}{2\kappa}{}^{\left(  3\right)  }R\right]
+\left(  2\kappa\right)  \left[  G_{ijkl}\pi^{ij}\pi^{kl}+\frac{\pi^{2}}%
{4}\right]  \frac{2\left(  f^{\prime}\left(  R\right)  -1\right)  }{f^{\prime
}\left(  R\right)  }\right.
\]%
\begin{equation}
\left.  +\frac{V(\mathcal{P})}{2\kappa f^{\prime}\left(  R\right)  }\right\}
=-\frac{\Lambda_{c}}{\kappa}\int_{\Sigma}d^{3}x\sqrt{g},\label{GWDW}%
\end{equation}
where we have assumed that $f^{\prime}\left(  R\right)  \neq0$ and we have
dropped a divergence form term. Eq.$\left(  \ref{GWDW}\right)  $ can be cast
in the form of Eq.$\left(  \ref{WDW2}\right)  $, by formally repeating the
same procedure. Thus one gets%
\[
\frac{1}{V}\frac{\left\langle \Psi\left\vert \int_{\Sigma}d^{3}x\left[
\hat{\Lambda}_{\Sigma}^{\left(  2\right)  }\right]  \right\vert \Psi
\right\rangle }{\left\langle \Psi|\Psi\right\rangle }+\frac{2\kappa}{V}%
\frac{2\left(  f^{\prime}\left(  R\right)  -1\right)  }{f^{\prime}\left(
R\right)  }\frac{\left\langle \Psi\left\vert \int_{\Sigma}d^{3}x\left[
G_{ijkl}\pi^{ij}\pi^{kl}+\pi^{2}/4\right]  \right\vert \Psi\right\rangle
}{\left\langle \Psi|\Psi\right\rangle }%
\]%
\begin{equation}
+\frac{1}{V}\frac{\left\langle \Psi\left\vert \int_{\Sigma}d^{3}%
xV(\mathcal{P})/\left(  2\kappa f^{\prime}\left(  R\right)  \right)
\right\vert \Psi\right\rangle }{\left\langle \Psi|\Psi\right\rangle }%
=-\frac{\Lambda_{c}}{\kappa}.\label{GWDW1}%
\end{equation}
From Eq.$\left(  \ref{GWDW1}\right)  $, we can define a \textquotedblleft%
\textit{modified}\textquotedblright\ $\hat{\Lambda}_{\Sigma}^{\left(
2\right)  }$ operator which includes $f^{\prime}\left(  R\right)  $. Thus, we
obtain%
\[
\frac{\left\langle \Psi\left\vert \int_{\Sigma}d^{3}x\left[  \hat{\Lambda
}_{\Sigma,f\left(  R\right)  }^{\left(  2\right)  }\right]  \right\vert
\Psi\right\rangle }{\left\langle \Psi|\Psi\right\rangle }+\frac{\kappa}%
{V}\frac{\left(  f^{\prime}\left(  R\right)  -1\right)  }{f^{\prime}\left(
R\right)  }\frac{\left\langle \Psi\left\vert \int_{\Sigma}d^{3}x\left[
\pi^{2}\right]  \right\vert \Psi\right\rangle }{\left\langle \Psi
|\Psi\right\rangle }%
\]%
\begin{equation}
+\frac{1}{V}\frac{\left\langle \Psi\left\vert \int_{\Sigma}d^{3}%
x\frac{V(\mathcal{P})}{2\kappa f^{\prime}\left(  R\right)  }\right\vert
\Psi\right\rangle }{\left\langle \Psi|\Psi\right\rangle }=-\frac{\Lambda_{c}%
}{\kappa},\label{GWDW2}%
\end{equation}
where%
\begin{equation}
\hat{\Lambda}_{\Sigma,f\left(  R\right)  }^{\left(  2\right)  }=\left(
2\kappa\right)  h\left(  R\right)  G_{ijkl}\pi^{ij}\pi^{kl}-\frac{\sqrt{g}%
}{2\kappa}\ ^{3}\!R^{lin},
\end{equation}
with%
\begin{equation}
h\left(  R\right)  =1+\frac{2\left[  f^{\prime}\left(  R\right)  -1\right]
}{f^{\prime}\left(  R\right)  }\label{h(R)}%
\end{equation}
and where $^{3}R^{lin}$ is the linearized scalar curvature. Note that when
$f\left(  R\right)  =R$, consistently it is $h\left(  R\right)  =1$. From
Eq.$\left(  \ref{GWDW2}\right)  $, we redefine $\Lambda_{c}$%
\begin{equation}
\Lambda_{c}^{\prime}=\Lambda_{c}+\frac{1}{2V}\frac{\left\langle \Psi\left\vert
\int_{\Sigma}d^{3}x\frac{V(\mathcal{P})}{f^{\prime}\left(  R\right)
}\right\vert \Psi\right\rangle }{\left\langle \Psi|\Psi\right\rangle }%
=\Lambda_{c}+\frac{1}{2V}\int_{\Sigma}d^{3}x\sqrt{g}\frac{Rf^{\prime}\left(
R\right)  -f\left(  R\right)  }{f^{\prime}\left(  R\right)  }%
,\label{NewLambda}%
\end{equation}
where we have explicitly used the definition of $V(\mathcal{P})$. In the same
spirit of the previous section, we find that by replacing $\Lambda_{0}\left(
\mu_{0},r\right)  $ with $\Lambda_{0}^{\prime}\left(  \mu_{0},r\right)  $, the
TT tensors one loop contribution for a $f\left(  R\right)  $ theory is given
by Eq.$\left(  \ref{lambda0mu}\right)  $ and the extremum is given therefore
by\footnote{For a complete derivation of the effective action for a $f\left(
R\right)  $ theory, see Ref.\cite{CENOZ}.}\footnote{By a canonical
decomposition of the gauge part $\xi_{a}$ into a transverse part $\xi_{a}^{T}$
with $\nabla^{a}\xi_{a}^{T}=0$ and a longitudinal part $\xi_{a}^{\parallel}$
with $\xi_{a}^{\parallel}=\nabla_{a}\psi$, it is possible to show that most of
the contribution comes from the longitudinal part (scalar). Evidence against
scalar perturbation contribution in a Schwarzschild background has been
discussed in Ref.\cite{Remo2}.}%
\begin{equation}
\Lambda_{0}^{\prime}\left(  \mu_{0},\bar{x}\right)  =\frac{G\mu_{0}^{4}}{2\pi
e^{2}},\label{Lambda0prime}%
\end{equation}
with $\bar{x}$ expressed by Eq.$\left(  \ref{sol}\right)  .$ In terms of
$\Lambda_{0}\left(  \mu_{0},\bar{x}\right)  $, we find%
\begin{equation}
\frac{1}{\sqrt{h\left(  R\right)  }}\left[  \Lambda_{0}\left(  \mu_{0},\bar
{x}\right)  +\frac{1}{2V}\int_{\Sigma}d^{3}x\sqrt{g}\frac{Rf^{\prime}\left(
R\right)  -f\left(  R\right)  }{f^{\prime}\left(  R\right)  }\right]
=\frac{G\mu_{0}^{4}}{2\pi e^{2}}\label{Lambda0primeex}%
\end{equation}
and isolating $\Lambda_{0}\left(  \mu_{0},\bar{x}\right)  $, we obtain%
\begin{equation}
\Lambda_{0}\left(  \mu_{0},\bar{x}\right)  =\sqrt{h\left(  R\right)  }%
\frac{G\mu_{0}^{4}}{2\pi e^{2}}-\frac{1}{2V}\int_{\Sigma}d^{3}x\sqrt{g}%
\frac{Rf^{\prime}\left(  R\right)  -f\left(  R\right)  }{f^{\prime}\left(
R\right)  }.
\end{equation}
Note that $\Lambda_{0}\left(  \mu_{0},\bar{x}\right)  $ can be set to zero
when%
\begin{equation}
\sqrt{h\left(  R\right)  }\frac{G\mu_{0}^{4}}{2\pi e^{2}}=\frac{1}{2V}%
\int_{\Sigma}d^{3}x\sqrt{g}\frac{Rf^{\prime}\left(  R\right)  -f\left(
R\right)  }{f^{\prime}\left(  R\right)  }.\label{lambda0_fin}%
\end{equation}
Let us see what happens when%
\begin{equation}
f\left(  R\right)  =\exp\left(  -\alpha R\right)  .\label{f(R)}%
\end{equation}
This choice is simply suggested by the regularity of the function at every
scale. In this case, Eq.$\left(  \ref{lambda0_fin}\right)  $ becomes%
\begin{equation}
\sqrt{\frac{3\alpha\exp\left(  -\alpha R\right)  +2}{\alpha\exp\left(  -\alpha
R\right)  }}\frac{G\mu_{0}^{4}}{\pi e^{2}}=\frac{1}{\alpha V}\int_{\Sigma
}d^{3}x\sqrt{g}\left(  1+\alpha R\right)  .
\end{equation}
For Schwarzschild, it is $R=0$, and by setting $\alpha=G$, we have the
relation%
\begin{equation}
\mu_{0}^{4}=\frac{\pi e^{2}}{G}\sqrt{\frac{1}{\left(  3G+2\right)  G}}.
\end{equation}
It is clear that the passage to the naked singularity is straightforward, at
least at formal level. The result is identical to Eq.$\left(
\ref{Lambda0prime}\right)  $ with the replacement%
\begin{equation}
\Lambda_{0}^{\prime}\left(  \mu_{0},\bar{x}\right)  \rightarrow\Lambda
_{0}^{\left(  naked\right)  \prime}\left(  \mu_{0},\bar{x}\right)  ,
\end{equation}
which means that also in Eq.$\left(  \ref{Lambda0primeex}\right)  $ we have to
replace $\Lambda_{0}\left(  \mu_{0},\bar{x}\right)  $ with $\Lambda
_{0}^{naked}\left(  \mu_{0},\bar{x}\right)  $. However, also in this case we
have the freedom to choose the r.h.s of Eq.$\left(  \ref{Lambda0primeex}%
\right)  $ in such a way to cancel $\Lambda_{0}^{naked}\left(  \mu_{0},\bar
{x}\right)  $. This means that also a naked Schwarzschild singularity predicts
a vanishing cosmological constant. Let us see the consequences on the
renormalization point $\mu_{0}$. If we further proceed and we fix the form of
$f\left(  R\right)  $ like in Eq.$\left(  \ref{f(R)}\right)  $, we get the
relation%
\begin{equation}
\mu_{0}^{\left(  naked\right)  4}=0.328\mu_{0}^{4}\qquad\longrightarrow
\qquad\mu_{0}^{naked}=0.757\mu_{0}.\label{mu0}%
\end{equation}
As we can see, it seems that a generic $f\left(  R\right)  $ theory cannot be
more different by the ordinary case, namely when $f\left(  R\right)  =R$. This
means that in this approach and at this level of approximation we cannot
discriminate the different geometries. The situation is more marked, if we had
chosen as a boundary condition for the naked singularity a privileged point
$r_{0}=2MG$, namely a \textquotedblleft\textit{fictitious throat}%
\textquotedblright. We use the term fictitious because there is no throat at
all. Nevertheless, if one wishes to fix such a boundary condition, there
should not be difference at all between positive and negative Schwarzschild
mass. Therefore, it appears that we are in a position where we cannot build a
spectrum of geometries \underline{\textit{including}} a naked singularity. On
the other hand, the impact of this approach on the cosmological constant
problem deserves further investigation.

\end{document}